\documentclass[12pt]{iopart}

\usepackage{iopams}

\newcommand{\dd}{\mbox{d}}

\begin{document}

\title[Path integral formulation of fBm for general Hurst exponent]
{Path integral formulation of fractional Brownian motion for general
Hurst exponent}

\author{I Calvo$^1$ and R S\'anchez$^2$}

\address{$^1$ Laboratorio Nacional de Fusi\'on, Asociaci\'on
EURATOM-CIEMAT, 28040 Madrid, Spain}

\address{$^2$ Fusion Energy Division, Oak Ridge National Laboratory,
Oak Ridge, TN 37831, U.S.A.}

\eads{\mailto{ivan.calvo@ciemat.es}, \mailto{sanchezferlr@ornl.gov}}


\begin{abstract}
In {\em J.~Phys.~A: Math.~Gen.} 28, 4305 (1995), K.~L. Sebastian gave
a path-integral computation of the propagator of subdiffusive
fractional Brownian motion (fBm), i.e. fBm with a Hurst or
self-similarity exponent $H\in(0,1/2)$. The extension of Sebastian's
calculation to superdiffusion, $H\in(1/2,1]$, becomes however quite
involved due to the appearance of additional boundary conditions on
fractional derivatives of the path. In this paper, we address the
construction of the path integral representation in a different
fashion, which allows to treat both subdiffusion and superdiffusion on
an equal footing. The derivation of the propagator of fBm for general
Hurst exponent is then performed in a neat and unified way.
\end{abstract}


\section{Introduction}

The Langevin representation for Brownian motion has been known for a
long time~\cite{Langevin1908,Uhlenbeck30}. In one dimension, it can be
formally written as
\begin{equation}\label{eq1}
x(t) = x_0 + \int_0^t \xi(t')\dd t',
\end{equation}
where $\xi(t)$ is a Gaussian, uncorrelated noise function:
\begin{equation}
\left\langle \xi(t)\xi(t')\right\rangle = \delta(t-t').
\end{equation}
The propagator of Brownian motion, which gives the probability of
finding the Brownian walker at some location $x$ at time $t>0$ after
having started at $x_0$ at time $0$, is a Gaussian whose width grows
as $t^{1/2}$:
\begin{equation}
G(x,t|x_0,0) =\frac{1}{\sqrt{2\pi t}}
\exp\left(-\frac{(x-x_0)^2}{2t}\right).
\end{equation}

However, many interesting processes (see \cite{Metzler2004} for a
review) ranging from hydrology~\cite{Molz97} to
finance~\cite{book:Elliott} or tracer transport by turbulent
flows~\cite{Sanchez06} exhibit anomalous diffusion, $\langle
(x-x_0)^2\rangle \propto t^{2H}$, with $H\neq 1/2$. In many situations
the anomalous behaviour is due to spatio-temporal correlations, which
translate into long-term time correlations in Lagrangian quantities. A
generalization of Brownian motion which was proposed to include such
correlations is known as fractional Brownian motion
(fBm)~\cite{Mandelbrot68}. Although slightly different representations
can be found in the literature, the one more convenient for our
purposes is
\begin{equation}\label{eq4}
x(t) = x_0 + \frac{1}{\Gamma(H+\frac{1}{2})}\int_0^t
(t-t')^{H-\frac{1}{2}}\xi(t')\dd t',
\end{equation}
where $\Gamma(\cdot)$ is the Gamma function and $\xi(t)$ is still
Gaussian and uncorrelated~\footnote{In \cite{Mandelbrot68,Taqqu}
fractional Brownian motion is defined by (\ref{eq4}) except that the
lower limit of the integral is taken at $-\infty$ and the issue of
boundary conditions is not present. Our definition is better suited
for the class of applications we are interested in (e.g. analysis of
series of Lagrangian velocities in turbulent plasmas). For such
physical problems, the stochastic differential equation is naturally
formulated in a domain $t\in(t_0,\infty)$, with finite $t_0$. An
equation analogous to (\ref{eq4}) was used in \cite{Lutz2001}, where
fBm and fractional time processes are compared.}. The exponent $H$
corresponds to both the self-similarity exponent of the
process~\cite{Mandelbrot68} and the Hurst exponent~\cite{Hurst51} of
the incremental process $\dd x(t):=x(t+\dd t)-x(t)$. On the one hand,
the average motion is invariant under the transformation $(x,t)
\rightarrow (\mu^H x, \mu t)$. On the other hand, the increments $\dd
x(t)$ are uncorrelated for $H=1/2$ and anticorrelated (correlated) for
$H\in(0,1/2)$ ($H\in(1/2,1]$). Consequently, (\ref{eq4}) can be used
to model diffusive ($H=1/2$), subdiffusive ($H\in(0,1/2)$), and
superdiffusive ($H\in(1/2,1]$) transport processes.

In a beautiful paper~\cite{Sebastian95}, K.~L. Sebastian showed that
it is very advantageous to rewrite fBm in terms of fractional
differential operators,
\begin{equation}\label{eq:fBmfracint}
x(t) = x_0 + {}_0D_t^{-(H+1/2)}\xi,
\end{equation}
where ${}_0D_t^{-\alpha}$ is the Riemann-Liouville fractional integral
of order $\alpha$~\cite{OldSpa,Podlubny}. Indeed, by combining path
integral methods~\cite{book:FeynmanHibbs} with fractional calculus, he
computed the exact propagator of (\ref{eq4}) for $H\le 1/2$. However,
the case $H>1/2$ is not addressed in \cite{Sebastian95} because the
problem becomes quite more complicated from the technical point of
view. The difficulties may be explained as follows. Sebastian
constructs the path integral by inverting (\ref{eq:fBmfracint}) in
terms of Riemann-Liouville fractional differential operators. This is
easy for $H\le 1/2$ because such inversion is well-defined simply
imposing the boundary conditions $x(0)=x_0$ and $x(T)=x_T$. However,
if $H>1/2$ additional boundary conditions involving fractional
derivatives at $t=0$ are required. Not only is the physical meaning of
these `fractional initial conditions' unclear, but also the
computation along the lines of \cite{Sebastian95} results too
cumbersome.

In this paper, we formulate the path integral for fBm using instead
$\xi(t)$ as the integration variable. This choice avoids the need to
invert (\ref{eq:fBmfracint}), and thus the need for the additional
fractional initial conditions. In our approach the difficulty resides
in how to manage the boundary conditions of $x(t)$, which translate
into a non-local constraint on the admissible realizations of the
noise. We will see below that the usage of Lagrange multipliers allows
to do this in an easy and transparent way, thus providing a general
path-integral derivation of the well-known propagator of
fBm~\cite{Wang90}.

\section{Path integral calculation of the propagator}
\label{sec2}

Given a Langevin equation describing the motion of a particle, the
propagator $G(x_T,T|x_0,0)$ is defined as the probability to find the
particle at $x=x_T$ at time $t=T$ if initially, $t=0$, it was located
at $x=x_0$. Hence, the propagator can be viewed as the motion of the
particle averaged over all realizations of the noise compatible with
the boundary conditions $x(0)=x_0$, $x(T)=x_T$. In the path integral
formalism this is formulated as follows. The essential object is the
probability measure ${\cal P}(\xi(t)){\cal D}\xi(t)$ on the space of
maps $\xi(t):[0,T]\to{\mathbb R}$. Since $\xi(t)$ is uncorrelated in
time and distributed as a Gaussian for each $t$, this is naturally
defined as
\begin{equation}
{\cal P}(\xi(t)){\cal D}\xi(t) = \exp\left(-\frac{1}{2}\int_0^T\xi(t)^2\dd
t\right){\cal D}\xi(t).
\end{equation}

We want to compute the propagator of (\ref{eq:fBmfracint}). Firstly,
note that the boundary conditions $x(0)=x_0$, $x(T)=x_T$ translate
into the following constraint on $\xi(t)$:
\begin{equation}\label{eq:constraintNoise}
{}_0D_T^{-(H+1/2)}\xi = x_T - x_0.
\end{equation}
Therefore, the propagator can be written as the expectation value:
\begin{equation}\label{eq:PathIntProp}
\fl G(x_T,T|x_0,0) = \int\delta\left({}_0D_T^{-(H+1/2)}\xi - (x_T - x_0)\right)
\exp\left(-\frac{1}{2}\int_0^T\xi(t)^2\dd t\right){\cal D}\xi(t),
\end{equation}
where the Dirac delta function ensures that we only integrate over maps
$\xi(t)$ satisfying (\ref{eq:constraintNoise}).

We proceed to compute the path integral on the right-hand side of
(\ref{eq:PathIntProp}). Define the action
\begin{equation}
S[\xi(t)] = \frac{1}{2}\int_0^T\xi(t)^2\dd t,
\end{equation}
and consider infinitesimal variations on the space of maps $\xi(t)$
satisfying (\ref{eq:constraintNoise}). We denote by $\bar\xi(t)$ the
map which makes the action stationary under such variations. Observe
that, in particular, $\bar\xi(t)$ verifies
(\ref{eq:constraintNoise}). Now, we perform the following change of
variables in (\ref{eq:PathIntProp}):
\begin{equation}
\xi(t) = \bar\xi(t) + \eta(t).
\end{equation}
The constraint (\ref{eq:constraintNoise}) implies that
\begin{equation}
{}_0D_T^{-(H+1/2)}\eta = 0.
\end{equation}

Using that the action is quadratic in $\xi(t)$ and that $\bar\xi(t)$
makes it stationary,
\begin{equation}\label{eq:PathIntProp2}
G(x_T,T|x_0,0) = e^{-S[\bar\xi(t)]}
\int\delta\left({}_0D_T^{-(H+1/2)}\eta\right)
e^{-S[\eta(t)]}{\cal D}\eta(t),
\end{equation}
whence we deduce that the propagator is of the form
\begin{equation}\label{eq:PathIntProp3}
G(x_T,T|x_0,0) = f(T)e^{-S[\bar\xi(t)]}\ .
\end{equation}
The function $f(T)$ will be determined at the end of the calculation
by imposing the normalization condition
\begin{equation}\label{eq:normalization}
\int_{-\infty}^\infty G(x_T,T|x_0,0)\dd x_T = 1,\ \forall T.
\end{equation}

It only remains to compute the stationary points of $S[\xi(t)]$
subject to the constraint (\ref{eq:constraintNoise}). It is difficult
to do it directly, for the condition (\ref{eq:constraintNoise}) is
hard to implement. The technique of Lagrange multipliers saves the
day, though. Our problem is equivalent to finding the stationary
points of
\begin{equation}
\tilde S[\xi(t),\lambda] = \frac{1}{2}\int_0^T\xi(t)^2\dd t +
\lambda\left({}_0D_T^{-(H+1/2)}\xi - (x_T - x_0)\right),
\end{equation}
under {\it arbitrary} infinitesimal variations of $\xi(t)$ and the
Lagrange multiplier $\lambda\in{\mathbb R}$.

From variations with respect to $\lambda$,
\begin{equation}
\delta_\lambda\tilde S[\xi(t),\lambda] =
\delta\lambda\left({}_0D_T^{-(H+1/2)}\xi - (x_T - x_0)\right),
\end{equation}
we recover, of course, the constraint (\ref{eq:constraintNoise}):
\begin{equation}\label{eq:ConstraintAgain}
{}_0D_T^{-(H+1/2)}\bar\xi = x_T-x_0.
\end{equation}

Let us perform now variations in $\xi(t)$:
\begin{equation}
\delta_\xi\tilde S[\xi(t),\lambda] = \int_0^T\xi(t)\delta\xi(t)\dd t
+ \lambda\ {}_0D_T^{-(H+1/2)}\delta\xi.
\end{equation}
Writing explicitly the fractional integral:
\begin{eqnarray}
\fl\delta_\xi\tilde S[\xi(t),\lambda] &=& \int_0^T\xi(t)\delta\xi(t)\dd t
+ \frac{\lambda}{\Gamma(H+1/2)}\int_0^T (T-t)^{H-1/2}\delta\xi(t)\dd t\\[5pt]
&=&\int_0^T\left(\xi(t) + \frac{\lambda}{\Gamma(H+1/2)}
(T-t)^{H-1/2}\right)\delta\xi(t)\dd t.
\end{eqnarray}
Since $\delta_\xi\tilde S[\bar\xi(t),\lambda]$ must vanish for
arbitrary $\delta\xi(t)$, we deduce
\begin{equation}\label{eq:StatSol}
\bar\xi(t) + \frac{\lambda}{\Gamma(H+1/2)}
(T-t)^{H-1/2} = 0.
\end{equation}

The Lagrange multiplier is determined by applying ${}_0D_T^{-(H+1/2)}$
to Eq.~(\ref{eq:StatSol}) and using (\ref{eq:ConstraintAgain}):
\begin{equation}\label{eq:StatSol2}
x_T-x_0 + \frac{\lambda}{\Gamma(H+1/2)^2}\int_0^T
(T-t)^{2H-1}\dd t = 0.
\end{equation}
Immediately, we find
\begin{equation}\label{eq:StatSol3}
x_T-x_0 + \frac{\lambda}{\Gamma^2(H+1/2)}\frac{T^{2H}}{2H} = 0,
\end{equation}
and
\begin{equation}\label{eq:StatSol4}
\lambda = - 2H\Gamma^2(H+1/2)\frac{x_T-x_0}{T^{2H}}.
\end{equation}

Inserting this in (\ref{eq:StatSol}):
\begin{equation}\label{eq:StatSolFinal}
\bar\xi(t) = 2H\Gamma(H+1/2)\frac{x_T-x_0}{T^{2H}}
(T-t)^{H-1/2}.
\end{equation}

Going back to (\ref{eq:PathIntProp3}) we straightforwardly obtain
\begin{equation}\label{eq:PathIntPropFinal}
\fl G(x_T,T|x_0,0) =
f(T)\exp\left(-H\Gamma^2(H+1/2)\frac{(x_T-x_0)^2}{T^{2H}}
\right).
\end{equation}
And $f(T)$ is obtained from the normalization property,
(\ref{eq:normalization}):
\begin{equation}
f(T) = \sqrt{\frac{H}{\pi }}\ \frac{\Gamma(H+1/2)}{T^{H}}.
\end{equation}
Hence, finally,
\begin{equation}\label{eq:PathIntPropFinal2}
\fl G(x_T,T|x_0,0) = \sqrt{\frac{H}{\pi }}\
\frac{\Gamma(H+1/2)}{T^{H}}\exp\left(-H\Gamma^2(H+1/2)
\frac{(x_T-x_0)^2}{T^{2H}} \right).
\end{equation}

\subsection{The kinetic equation}

Once the propagator is known, we can easily derive the kinetic
equation associated to fBm. The Fourier transform of $G(x,t|x_0,0)$
with respect to $x$ is
\begin{equation}\label{eq66b}
\hat G(k,t)  = \exp\left(-
\frac{t^{2H}}{4H\Gamma^2\left(H+1/2\right)} k^2
\right).
\end{equation}
Differentiating with respect to $t$:
\begin{equation}\label{eq67b}
\partial_t \hat G(k,t) =
- \frac{t^{2H-1}}{2\Gamma^2\left(H+1/2\right)}k^2
\hat G(k,t).
\end{equation}
Finally, we Fourier invert this expression to get the kinetic
equation:
\begin{equation}\label{eq68b}
\partial_t G (x,t) =
\frac{t^{2H-1}}{2\Gamma^2\left(H+1/2\right)}
\partial^2_x G(x,t),
\end{equation}
which is a diffusion equation with time-dependent diffusivity, as
originally deduced in \cite{Wang90} from arguments based on the
fluctuation-dissipation theorem. The invariance of (\ref{eq68b}) under
the transformation $(x,t) \rightarrow (\mu^H x, \mu t)$ is
manifest. Obviously, when $H=1/2$ we retrieve the standard diffusion
equation associated to Brownian motion:
\begin{equation}
\partial_t G(x,t) =  \frac{1}{2}\partial^2_x G(x,t).
\end{equation}

\section{Conclusions}

In this paper we have computed by path integral methods the well-known
 propagator of fBm for general Hurst exponent, extending the
 construction of \cite{Sebastian95}, which was restricted to the
 subdiffusive case. The key technical point in our approach is to use
 the noise as path integral variable instead of the coordinate of the
 particle trajectory.

\vskip 0.2cm

{\bf Acknowledgements:} I. C. acknowledges the hospitality of Oak
Ridge National Laboratory, were this work was carried out. Part of
this research was sponsored by the Laboratory Research and Development
Program of Oak Ridge National Laboratory, managed by UT-Battelle, LLC,
for the US Department of Energy under contract number
DE-AC05-00OR22725.


\section*{References}

\begin{thebibliography}{10}

\bibitem{Langevin1908}
P.~Langevin.
\newblock {\em C. R. Acad. Sci. Paris}, 146:530--533, 1908.

\bibitem{Uhlenbeck30}
G.~E. Uhlenbeck and L.~S. Ornstein.
\newblock {\em Phys. Rev.}, 36:823--841, 1930.

\bibitem{Metzler2004}
R. Metzler and J. Klafter.
\newblock {\em J. Phys. A: Math. Gen.}, 37:R161-R208, 2004.

\bibitem{Molz97}
F.~J. Molz, H.~H. Liu, and J.~Szulga.
\newblock {\em Water Resour. Res.}, 33:2273, 1997.

\bibitem{book:Elliott}
R.~J. Elliott and J.~van~der Hoek.
\newblock {\em {Fractional Brownian motion and financial modeling}}.
\newblock In Trends in Mathematics, Birkh\"auser Verlag, Basel, 2001.

\bibitem{Sanchez06}
R.~S\'anchez, B.~A. Carreras, D.~E. Newman, V.~E. Lynch, and B.~Ph. van
  Milligen.
\newblock {\em Phys. Rev. E}, 74:016305, 2006.

\bibitem{Mandelbrot68}
B.~B. Mandelbrot and J.~W. van Ness.
\newblock {\em SIAM Rev.}, 10:422, 1968.

\bibitem{Taqqu}
G.~Samorodnitsky and M.~S. Taqqu.
\newblock {\em {Stable non-Gaussian processes}}.
\newblock Chapman \& Hall, New York, 1994.

\bibitem{Lutz2001}
E. Lutz.
\newblock {\em Phys. Rev. E}, 64:051106, 2001.

\bibitem{Hurst51}
H.~E. Hurst.
\newblock {\em Trans. Am. Soc. Civ. Eng.}, 116:770, 1951.

\bibitem{Sebastian95}
K.~L. Sebastian.
\newblock {\em J. Phys. A: Math. Gen.}, 28:4305, 1995.

\bibitem{OldSpa}
K.~Oldham and J.~Spanier.
\newblock {\em {The Fractional Calculus}}.
\newblock Academic Press, New York, 1974.

\bibitem{Podlubny}
I.~Podlubny.
\newblock {\em {Fractional differential equations}}.
\newblock Academic Press, New York, 1998.

\bibitem{book:FeynmanHibbs}
R.~P. Feynman and A.~R. Hibbs.
\newblock {\em {Quantum Mechanics and Path Integrals}}.
\newblock McGraw-Hill, New York, 1965.

\bibitem{Wang90}
K.~G. Wang and C.~W. Lung.
\newblock {\em Phys. Lett. A}, 151:119, 1990.

\end{thebibliography}
\end{document}